\documentclass[iop]{emulateapj}

\begin{document}

\title{Magnification by Galaxy Group Dark Matter Halos}
\author{
Jes Ford\altaffilmark{1}, 
Hendrik Hildebrandt\altaffilmark{1,2}, 
Ludovic Van Waerbeke\altaffilmark{1}, 
Alexie Leauthaud\altaffilmark{3}, 
Peter Capak\altaffilmark{4}, 
Alexis Finoguenov\altaffilmark{5,6}, 
Masayuki Tanaka\altaffilmark{3}, 
Matthew R. George\altaffilmark{7,8}, 
Jason Rhodes\altaffilmark{9,10}
}

\altaffiltext{1}{Department of Physics and Astronomy, University of British Columbia, 6224 Agricultural Road, Vancouver, BC V6T 1Z1, Canada}
\altaffiltext{2}{Argelander-Institut f\"ur Astronomie, Auf dem H\"ugel 71, 53121 Bonn, Germany}
\altaffiltext{3}{Institute for the Physics and Mathematics of the Universe, The University of Tokyo, Chiba 277-8582, Japan}
\altaffiltext{4}{NASA Spitzer Science Center, California Institute of Technology, 220-6 Caltech, 1200 East California Boulevard, Pasadena, CA 91125, USA}
\altaffiltext{5}{Max-Planck-Institut fuer Extraterrestrische Physik, Giessenbachstrasse 1, D-85748 Garching, Germany}
\altaffiltext{6}{Center for Space Science Technology, University of Maryland Baltimore County, 1000 Hilltop Circle, Baltimore, MD 21250, USA}
\altaffiltext{7}{Department of Astronomy, University of California, Berkeley, CA 94720, USA}
\altaffiltext{8}{Lawrence Berkeley National Laboratory, 1 Cyclotron Road, Berkeley, CA 94720, USA}
\altaffiltext{9}{Jet Propulsion Laboratory, California Institute of Technology, Pasadena, CA 91109, USA}
\altaffiltext{10}{California Institute of Technology, 1200 East California Boulevard, Pasadena, CA 91125, USA}

\begin{abstract}
We report on the detection of gravitational lensing magnification by a population of galaxy groups, at a significance level of 4.9$\sigma$.  Using X-ray selected groups in the COSMOS 1.64 deg$^2$ field, and high-redshift Lyman break galaxies as sources, we measure a lensing-induced angular cross-correlation between the samples.  After satisfying consistency checks that demonstrate we have indeed detected a magnification signal, and are not suffering from contamination by physical overlap of samples, we proceed to implement an optimally weighted cross-correlation function to further boost the signal to noise of the measurement. Interpreting this optimally weighted measurement allows us to study properties of the lensing groups. We model the full distribution of group masses using a composite-halo approach, considering both the singular isothermal sphere and Navarro-Frenk-White profiles, and find our best fit values to be consistent with those recovered using the weak-lensing shear technique.  We argue that future weak-lensing studies will need to incorporate magnification along with shear, both to reduce residual systematics and to make full use of all available source information, in an effort to maximize scientific yield of the observations.
\end{abstract}

\keywords{galaxies: photometry}                                                 
 
\setcounter{section}{0}
\setcounter{subsection}{0}
\setcounter{subsubsection}{0}

\section{Introduction}
Weak gravitational lensing is a unique tool for probing the mass distribution of the universe and for constraining dark matter halo properties of galaxies and clusters. In contrast to alternative mass estimate methods (employing e.g. X-ray temperatures, radial velocities, or mass-to-light ratios), weak lensing does not rely on any assumptions about virial equilibrium and is sensitive to all mass along the line of sight, making no distinction between luminous and dark matter.

Over the past decade, an enormous international effort has been invested in improving the reliability of weak lensing analysis \citep{step1, step2, great08, great10}, seeking to remove biases and systematic effects that limit the accuracy of the method. By far most of the work has been focused on measuring the shear signal, the coherent stretching and distortion of distant galaxy shapes by a foreground lensing mass, but recently the magnification signal has begun to attract attention as well \citep{Scranton05, Hildebrandt09b, Hildebrandt11, LHJM10, Umetsu11, Huff11}.  

Weak lensing magnification is, to first order, a measure of the convergence of a lensing mass. It can be detected through the stretching of solid angle on the sky, which leads to the amplification of source flux, since lensing conserves surface brightness (i.e. photons are neither created nor destroyed in purely lensing processes). In general, two different approaches can be taken to measure magnification. The method we employ here involves observing the effects on source number densities; an interesting alternative method is being explored by \citet{Schmidt12}, which makes use of source size and flux information, and employs the same COSMOS X-ray groups used in this study.

Magnification affects the source number densities in two ways, and the one that dominates is determined by the intrinsic magnitude number counts of the sources in question. Simply put, the brightest sources, which usually have steep number counts, will exhibit an {\it increase} in number density when lensed, as the amplification allows more objects to be detected, while the number density of the faintest sources, having relatively shallow number counts, will {\it decrease} \citep{Narayan89}.

Compared to shear measurements, magnification exhibits a slightly lower signal-to-noise ratio (S/N), the reason it has been largely ignored until recently. However, what magnification lacks in signal strength, it makes up for in terms of its ability to be applied to lenses at higher redshift and to poorly resolved sources \citep{Waerbeke10}. Since shear studies require measurements of galaxy shapes, in order for a source to be used it must necessarily be well resolved. This is in stark contrast to magnification studies using source number densities, which have no such requirement for the sources to be resolved at all! In principle, only source magnitudes, redshifts, and positions relative to a lens must be known. This simple fact makes it possible to extend weak-lensing magnification analyses to a much higher redshift than possible for shear, and allows a much higher source density to be included in the analysis.  See \citet{Waerbeke10}, \citet{RozoSchmidt10}, and \citet{Umetsu11} for more detailed discussions of the benefits of combining magnification with shear in gravitational lensing studies. 

In Section \ref{theory}, we review the equations describing the effects of weak-lensing magnification on source number densities. Section \ref{data} gives the properties of the X-ray groups and Lyman break galaxies (LBGs) that are used in this study. Then Section \ref{results} describes the steps of our analysis, and results of the composite-halo model fitting. We summarize the results in Section \ref{summary}, and compare with weak-lensing shear measurements that have previously been made on populations of galaxy groups. We use the WMAP7 $\Lambda$CDM cosmological parameters $H_0 =$ 71 km s$^{-1}$ Mpc$^{-1}$ and $\Omega_{\Lambda} = 0.734$ \citep{WMAP7}, and set $\Omega_M = 1 - \Omega_{\Lambda}$.

\section{Theory}
\label{theory}
The amplification matrix $\cal{A}$ maps the image deformation from the source to observer frame, and describes the first order effects of gravitational lensing:  
\begin{equation}
\cal{A} = \left( \begin{array}{cc}
{1-\kappa-\gamma_1} & {-\gamma_2} \\
{-\gamma_2} & {1-\kappa+\gamma_1} \\
\end{array} \right).
\end{equation}
It is a function of the convergence $\kappa$, and the shear $\gamma$, which define the isotropic and anisotropic focusing of light rays, respectively. The magnification factor $\mu$ is the inverse determinant of this matrix, so that
\begin{equation}
\mu = \frac{1}{\mathrm{det} \cal{A}} = 
\frac{1}{(1-\kappa)^2 - \left|\gamma\right|^2}
\end{equation}
\citep{BartelmannSchneider01}.  

The cumulative number counts of distant {\it unlensed} sources $N_0$ are related to the observed {\it lensed} number counts $N$, up to some flux $f$, by the equation
\begin{equation}
N (>f) = \frac{1}{\mu} N_0 \left( > \frac{f}{\mu} \right).
\end{equation}
Here the two distinct effects of weak-lensing magnification, on source number counts, are made explicit. The prefactor of $1 / \mu$ is the dilution of source density, as the observed solid angle on the sky is stretched by a foreground massive lens. The modification to the flux $f / \mu$ inside the argument of $N_0$ represents the effect of source amplification by a lens, such that one is able to detect intrinsically fainter objects due to gravitational lensing.

Switching from working in fluxes to magnitudes $m$, the differential number count relationship was demonstrated by \citet{Narayan89} to be
\begin{equation}
n(m)dm = \mu ^{\alpha -1} n_0 (m)dm,
\end{equation}
where $\alpha$ is defined according to
\begin{equation}
\alpha \equiv \alpha(m) = 2.5 \frac{d}{dm} \log n_0(m).
\end{equation}
Thus, distant source galaxies, lensed by an intervening concentration of mass, may have their observed number counts increased {\it or} decreased depending on the sign of the quantity $(\alpha -1)$. Sources for which $(\alpha -1) > 0$ will appear to be correlated on the sky with a lens position, while sources with $(\alpha -1) < 0$ will be anti-correlated, as a dearth of objects will be observed in the vicinity of a lens. The number density of galaxies for which the intrinsic number count slope gives $(\alpha -1) \approx 0$ will essentially be unaffected by lensing magnification, as the dilution and amplification effects will cancel, and no correlation signal will be observed for these objects \citep{Scranton05}.

\section{Data}
\label{data}
\subsection{Lenses}
The lenses in this study consist of X-ray-selected galaxy groups in the COSMOS field. See \citet{Leauthaud10} for the detailed properties of these groups. From the full sample of 206 groups investigated in the aforementioned study, we use the shear-calibrated mass estimates to construct the most massive subsample of groups for this magnification study. Here masses are characterized by the parameter $M_{200}$, the total mass interior to a sphere of radius $R_{200}$, within which the average density is 200 times the critical. 

\begin{figure*}
\begin{center}
\includegraphics[scale=0.7]{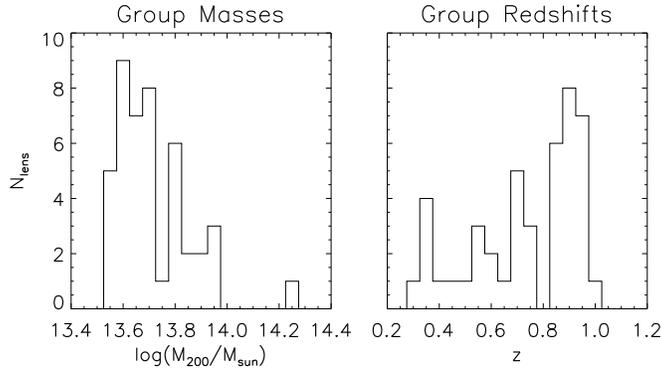}

\caption{Masses and photometric redshifts of the groups in this study. We select the most massive groups in our sample, $M_{200} / M_\odot \ge 3.56 \times 10^{13}$. Using only the cleanest groups (characterized by having $\ge$ 4 members, well-defined centroids, and no flags on possible mergers or projection effects), and applying appropriate masking, we are left with a sample of $44$ groups for this lensing magnification analysis.}
\label{hists}
\end{center}
\end{figure*}

Any groups that have less than four member galaxies, that appear to be undergoing mergers, that have uncertain centroids, or that raise concerns about projection effects, are excluded from the analysis. These restrictions follow from the group catalog requirement \texttt{FLAG\_INCLUDE=1}, discussed in \citet{George11}. The remaining $44$ most massive groups have shear determined masses in the range $ 3.56 \times 10^{13} \le M_{200} / M_\odot \le 1.70 \times 10^{14} $, and we employ stacking to increase the S/N of the magnification measurement. The redshift range of the groups is $ 0.32 \le z \le 0.98 $. Figure \ref{hists} displays these lens properties.

Choosing an optimal lens centroid about which to construct angular bins is an area of ongoing research, and common choices include the brightest central galaxy or the X-ray emission peak. If the location of the dark matter density peak were known a priori, then it would obviously be the ideal choice, but instead we must rely upon some combination of observables to approximate this position. In this paper, we define lensing mass centers by the location of the group galaxy with the highest stellar mass (MMGG$_\mathrm{scale}$) lying within a distance $ (R_s + \sigma_x) $ of the X-ray center, where $ R_s $ is the group scale radius and $ \sigma_x $ is the uncertainty in the X-ray center position \citep{George11}.  In order to be very confident about the locations of group centers, we exclude groups for which this galaxy is not the most massive member of the group. This choice of centroid has been shown to accurately trace the centers of halos in this sample by optimizing the shear signal on small scales \citep{George12}.

\subsection{Sources}
Background sources are LBGs, a type of high-redshift star-forming galaxy that has been used successfully in previous magnification studies \citep[see][]{Hildebrandt09b, Hildebrandt11}. These LBGs were selected using the typical three-color dropout technique. For the $U$-, $G$-, and $R$-dropouts the selections described in \citet{Hildebrandt09a} were used, however the COSMOS Subaru $g^+$ and $r^+$ data were used instead of the CFHT-LS $g^*$ and $r^*$ data (see \citet{Capak07} for the filter definitions). For the $B$-dropouts the selection from \citet{Ouchi04} was used.  

The appeal of using LBGs for magnification is rooted in the fact that their luminosity functions (LFs) have been extensively studied and their redshift distributions are fairly narrow and accurate. After all quality cuts and image masking, we are left with 45,132 LBGs in total. The four distinct sets are comprised of 12,980 $U$-, 22,520 $G$-, 4870 $B$-, and 4762 $R$-dropouts, located at redshifts of $\sim$3.1, 3.8, 4.0, and 4.8, respectively.

We first test our data selection by cross-correlating the foreground groups with LBGs separated into discrete magnitude bins. Here we use the basic \citet{LandySzalay93} estimator, 
\begin{equation}
w(\theta)=\frac{\text{D}_1 \text{D}_2 - \text{D}_1 \text{R} - \text{D}_2 \text{R} + \text{RR}}{\text{RR}},
\end{equation}
to simply compute cross-correlations between groups and background sources. D$_1$ and D$_2$ represent the data sets of lenses and sources, and R are the {\it random objects} from a mock catalog we create, containing points uniformly distributed throughout the COSMOS survey area. Each product of terms is the number of pairs of those objects found to lie within some angular bin, normalized by the total number of pairs found at all angular separations. This cross-correlation estimator has been shown to be both robust and unbiased \citep{Kerscher00}.  

In any lensing study, care must be taken to ensure that regions of an image containing artifacts such as saturated pixels, satellite tracks, or other spurious effects are masked out of the investigation. We consistently apply the same masks to the group, source, and random catalogs, prior to the correlation analysis. Using a large number (584,586) of objects in this random catalog serves to reduce shot noise.

\begin{figure*}
\begin{center}
\includegraphics[scale=0.9]{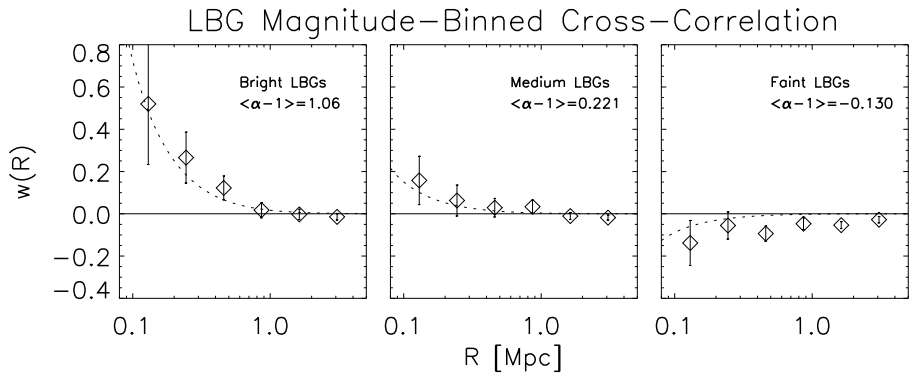}
\caption{Angular cross-correlation of the X-ray groups with Lyman break galaxies, the latter separated into three magnitude-selected samples. The bright sample contains $U$, $G$, $B$, and $R$-dropouts in the magnitude ranges $23<r<25$, $23.5<i<25$,  $23.5<i<25$,  and $24<z<25.5$, respectively. Similarly, the medium ranges are $25<r<25.5$, $25<i<26$,  $25<i<26$,  $25.5<z<26$. The faint ranges are $r>25.5$, $i>26$,  $i>26$,  $z>26$. These magnitude ranges are selected to contain LBGs for which $(\alpha-1)>0$, $\approx 0$, and $<0$. The measured correlations for each LBG sample are simply averaged here (weighting by the number counts) in order to more clearly display this diagnostic check. The dashed curves are calculated from the composite-NFW fit, using weighting by the appropriate $\langle \alpha -1 \rangle$ factor, which is given in each panel. The negative correlation observed for the faintest sample is a good indication that no redshift overlap exists between foreground lenses and background sources.}
\label{MagBinned}
\end{center}
\end{figure*}

We expect that the faintest (brightest) magnitude bins should yield a negative (positive) cross-correlation with the group centers, and this is exactly what we find. Figure \ref{MagBinned} displays this anticipated result, where we simply use a number count weighted average to combine the signal of the distinct LBG samples. As discussed in \citet{Hildebrandt09b}, this negative correlation is one of the strongest verifications that no redshift overlap exists between lens and source populations, for no viable reason other than lensing magnification can be given for such a signal to exist. Redshift overlap between samples must be avoided in magnification studies, as positive cross-correlations due to physical clustering would overwhelm any lensing-induced signal.

\section{Analysis and Results}
\label{results}
\subsection{Measuring $\alpha(m)$}
Along with the mass of the lens itself, the slope of the source number counts as a function of magnitude, parameterized by the quantity $\alpha \equiv \alpha(m)$, controls the amplitude and sign of the expected magnification signal. To interpret the correlations that we measure, and to implement an optimally weighted procedure, we must determine the value of this quantity for every source galaxy that we intend to use in the measurement. Fortunately, LBGs have been extensively studied and many measurements of their LFs have been published.  

For the $U$-, $G$-, and $R$-dropouts, we use the recent measurements by \citet{vanderBurg10}. For the $B$-dropouts we use the results of \citet{Sawicki06}. These two sets of measurements both involved fitting a Schechter function \citep{Schechter76} to their galaxy number counts, and their best-fit parameters that we use here are displayed in Table 1. The Schechter Function is given by
\begin{equation}
\Phi(M)=0.4\ln(10)\Phi^\ast10^{0.4(\alpha_{\text{LF}}+1)(M^\ast-M)} \exp [-10^{0.4(M^\ast-M)}],
\end{equation}
where $\Phi^\ast$, $M^\ast$, and $\alpha_{\text{LF}}$ are the normalization, characteristic magnitude, and faint-end slope of the LF. Note that the $\alpha(m)$ which we want to calculate is not the same as the constant parameter $\alpha_{\text{LF}}$, but approaches it in the limit of very faint magnitudes.

Solving this equation for $\alpha(m)$, we obtain
\begin{equation}
\alpha(m) = 2.5 \frac{d}{dm} \log n_0(m)= 2.5 \frac{d}{dM} \log \Phi(M)\]
\[ = 10^{0.4(M^\ast-M)}-\alpha_{\text{LF}}-1.
\end{equation}
We convert the observed apparent magnitudes $m$ of the LBGs to absolute magnitudes $M$ via the relationship $M = m - $DM$ + 2.5 \log (1+z)$, where DM and $z$ are the distance modulus and redshift of the galaxy in question. Since we select apparent magnitudes in the $r$, $i$, and $z$ bands for the $U$-, $G$- and $B$-, and $R$-dropouts, we probe very similar restframe wavelengths and the $K$-correction between the samples is negligible. Thus we ignore it here. Using the LF parameters in Table \ref{LFtable}, combined with the conversion to absolute magnitudes, we then obtain a measure of $\alpha(m)$ for every LBG in the sample.

It is important to assess how uncertainties in the quantity $\alpha(m)$ can affect the interpretation of the magnification measurement. Since we will rely on the quantity $(\alpha-1)$ as a weight factor in this analysis, using a wrong $\alpha$ could potentially lead to a bias in the mass measurement. For very faint objects the observed magnitudes become less certain due to shot noise. We propagate these magnitude errors through Equation 8 to obtain an uncertainty on $\alpha(m)$, which is used to find a magnitude-based cut on the sources. We find that cutting $\sim$10\% of the very faintest sources, largely $R$-dropouts, gives us a good balance between removing the most uncertain $\alpha$ values, but still retaining a significant number of sources for the analysis.

Here we consider two possible sources of systematic error, which are combined in quadrature to yield the total systematic error, reported in Section 4.3. The first source is uncertainty on the LF parameters (see Table 1), which includes the effects of cosmic variance. We repeat the composite-halo fit, detailed in Section 4.3, for the range of permitted values of $M^\ast$ and $\alpha_{\text{LF}}$, finding a maximum variation in the mass measurement of up to $\sim$40\%. Second, we consider the possibility for a small photometric offset to exist between the various surveys used in this work. Assuming a maximum offset of $\pm$0.05 magnitudes between surveys, we vary all observed source magnitudes uniformly by offsets in the range $-0.5 \leq \delta m \leq 0.5$, and find the maximum effect on the mass measurement to be $\sim$15\%.

\subsection{Optimally Weighted Cross-correlation}
We implement a modified version of the \citet{LandySzalay93} estimator for the angular cross-correlation function, in which pair counts are weighted by their expectations from the differential source number counts as a function of magnitude. This weighted correlation function has been shown to optimally boost the magnification signal \citep{Menard03}:    
\begin{equation}
w(\theta)_{\text{optimal}}=\frac{\text{S}^{\alpha-1} \text{L} - \text{S}^{\alpha-1} \text{R} - \langle \alpha-1 \rangle \text{LR}}{\text{RR}} + \langle \alpha-1 \rangle .
\end{equation}
Optimal-weighting was first implemented by \citet{Scranton05} and, apart from notation, this equation is identical to the estimator used in \citet{Hildebrandt09b}.  As with the original basic estimator, each term represents the number of pairs of objects found in a given angular $\theta$ bin, normalized by the total number of pairs at all angular separations.  

S stands for the {\it sources}, or background lensed galaxies, L are the {\it lenses}, or X-ray groups, and once again R are the {\it random objects}. The superscript $(\alpha-1)$ on the S indicates that pair counts involving sources are to be weighted by this factor. After removing masked objects from the catalogs, and satisfying the above selection criteria, we are left with 39,710 LBG sources, $44$ X-ray group lenses, and 584,586 random objects for the analysis.

\begin{table}
  \begin{center}

   \caption{Luminosity Function (Schechter) Parameters from External LBG Measurements. $^a$ LF parameters from \protect \citet{vanderBurg10}. $^b$ LF parameters from \protect \citet{Sawicki06}.}
 \label{LFtable}
    \begin{tabular}{lcccl}
      \hline \hline
      LBG Sample & $M^*$ & $\alpha_{\text{LF}}$ & Number \\ \hline
      $U$ ($z \sim$ 3.1)$^a$ & $-20.84^{+0.15}_{-0.13}$ & $-1.60^{+0.14}_{-0.11}$ & 12,980 \\
      $G$ ($z \sim$ 3.8)$^a$ & $-20.84^{+0.09}_{-0.09}$ & $-1.56^{+0.08}_{-0.08}$ & 22,520 \\
      $B$ ($z \sim$ 4.0)$^b$ & $-21.00^{+0.40}_{-0.46}$ & $-1.26^{+0.40}_{-0.36}$ & 4,870 \\ 
      $R$ ($z \sim$ 4.8)$^a$ & $-20.94^{+0.10}_{-0.11}$ & $-1.65^{+0.09}_{-0.08}$ & 4,762 \\
      \hline
    \end{tabular}
  \end{center}
\end{table}

The brightest source galaxies, which are observationally found to lie in the steepest part of the LF, are expected to be positively correlated with the group centers, have the largest value of $(\alpha-1)$, and so receive a relatively large weight in this correlation study. In contrast, the faintest background galaxies are expected to be anti-correlated, on average, with the group positions, because the effects of magnification dilution should be greater than the amplification of flux can compensate for, and these galaxies thus receive a negative weight. Sources for which $(\alpha-1) \approx 0$ ought to have the effects of dilution and amplification cancel out overall, and receive very little to no weight in this analysis \citep{Scranton05}. 

The optimally weighted correlation function is given in Figure 3, and shows the measured radial profiles for this stack of massive galaxy groups. Error bars are $1 \sigma$ uncertainties, obtained by jackknife resampling of the source population. To do this, we create 50 jackknife samples of data, each with a different $1/50$ of sources removed from it. Then we measure the optimal correlation function for each, and from these estimate the covariance matrix through
\begin{equation}
C(\theta_1, \theta_2)= \left( \frac{N}{N-1} \right)^2 \times \sum_{j=1}^N [w_j(\theta_1)-\bar{w}(\theta_1)] \times [w_j(\theta_2)-\bar{w}(\theta_2)],
\end{equation}
where the index $j$ runs over the $N=50$ jackknife measurements.

\subsection{Halo Mass Profiles}
\begin{figure*}
\begin{center}
\includegraphics[scale=0.75]{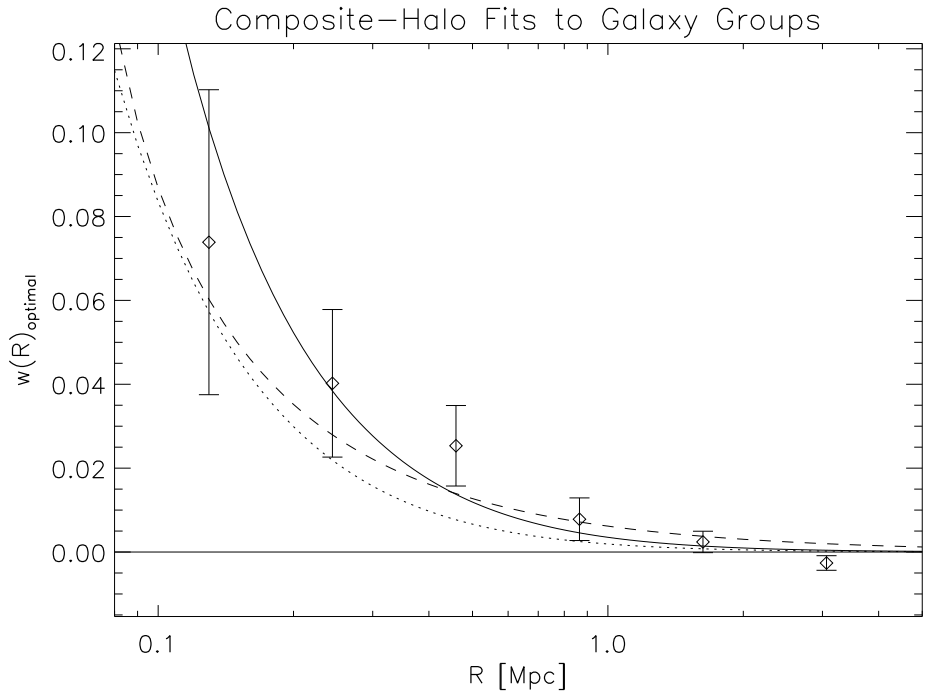}
\caption{Composite-halo fits to the optimally weighted correlation function, using the LBG background source sample. The significance of the magnification detection is 4.9$\sigma$. The dashed line is the composite-SIS and the solid line is the composite-NFW.  We find the best-fit relative scaling relations for each to be $a= M_{\text{mag}}/M_{\text{shear}}= 1.2 \pm 0.4 \pm 0.4^{\text{sys}}$ (SIS) and $a= 1.8 \pm 0.5 \pm 0.4^{\text{sys}}$ (NFW). The dotted line shows the prediction from the shear measured values of $M_{200}$ (A. Leauthaud 2011, private communication).}
\label{multihalo}
\end{center}
\end{figure*}

Measuring the magnification-induced effects on source number counts behind massive lenses allows one to estimate properties of the lens, such as the mass profile. In this paper, we use a composite-halo approach which allows us to fit for the full range of both group masses and redshifts. The horizontal axes in Figures \ref{MagBinned} and \ref{multihalo} are therefore actual transverse distances obtained by taking account of the angular diameter size at each unique group redshift. We incorporate both the singular isothermal sphere (SIS) and the Navarro-Frenk-White (NFW; \citep{nfw97}) density profiles into the composite-halo modeling.



The magnification contrast is $\delta\mu(\theta) \equiv \mu(\theta) -1$, and for an SIS halo it is simply given by 
\begin{equation}
\delta \mu_{\text{SIS}}(\theta)=\frac{\theta_{\text{E}}}{\theta-\theta_{\text{E}}},
\end{equation}
where $\theta_{\text{E}}=4\pi(\frac{\sigma_v}{c})^2\frac{D_{ls}}{D_s}$ is the Einstein radius of the lens, and $D_{ls}$ and $D_s$ are angular diameter distances between lens and source, and observer and source, respectively. The velocity dispersion of the lens, $\sigma_v$, can be expressed in terms of the mass and critical energy density of the universe at lens redshift $z$:
\begin{equation}
\sigma_v=\left[ \frac{\pi}{6}200\rho_{\text{crit}}(z)M_{200}^2 G^3 \right]^\frac{1}{6} .
\end{equation}

For the NFW halo, the magnification contrast takes a slightly more complicated form. From Equation 2, we have
\begin{equation}
\delta\mu_{\text{NFW}}(\theta)=\left[ (1-\kappa_{\text{NFW}})^2 - \left|\gamma_{\text{NFW}}\right|^2 \right]^{-1} -1.
\end{equation}
We use the analytical NFW expressions for $\kappa$ and $\gamma$ derived in \citet{WrightBrainerd00} to evaluate $\delta\mu_{\text{NFW}}$ for every lens-source pair in the study. The two NFW fit parameters are the scale radius $r_{\text{s}}$ and the concentration $c$, which together determine the mass
\begin{equation}
M_{200}=\frac{4\pi}{3}(200)\rho_{\text{crit}}(z)c^3r_{\text{s}}^3 .
\end{equation}
As we do not find this magnification measurement precise enough to provide meaningful two-parameter constraints, we use the mass-concentration relation given in \citet{Munoz11} to reduce this to a single-parameter fit.

We perform a composite-halo fit for both lens models, similar to the multi-SIS used in \citet{Hildebrandt11}. This allows us to fit for a range of masses and redshifts, thereby avoiding any biases that would be introduced by simply fitting to a stacked average lens profile. The optimally weighted correlation function is related to the magnification contrast through
\begin{equation}
w(a)_{\text{optimal}}= \frac{1}{N_{l}}\sum_{i=1}^{N_{l}} \langle (\alpha -1)^2 \rangle_i \delta \mu(z_i, aM_{\text{shear},i}),
\end{equation}
where $i$ runs over all lenses. Here the fit parameter $a$ characterizes the scaling relation between the $M_{200}$ previously measured from the shear, and the best fit $M_{200}$ from magnification, so that $a \equiv M_{\text{magnification}}/M_{\text{shear}}$. 

We use the generalized minimum-$\chi^2$ method to fit the composite-halo profiles to the magnification measurements (see Figure \ref{multihalo}), using the full unbiased inverse covariance matrix, according to the prescription in \citet{Hartlap07}. The $\chi^2/$dof is 1.5 for the composite-SIS and 0.8 for the composite-NFW ($\chi_{\text{SIS}}^2=7.5$, $\chi_{\text{NFW}}^2=4.2$, dof=5 in both cases). For the composite-SIS, we obtain a best-fit value of $a=1.2 \pm 0.4 \pm 0.4^{\text{sys}}$, and with the composite-NFW we get $a=1.8 \pm 0.5 \pm 0.4^{\text{sys}}$. These results indicate consistency with the previous shear mass measurements, albeit with large uncertainties.




\section{Summary and Conclusions}
\label{summary}
We report a 4.9$\sigma$ detection of weak-lensing magnification from a population of X-ray-selected galaxy groups. This is the first magnification measurement using source number densities successfully performed on the group scale. \citet{Schmidt12} have recently explored the magnification of these groups using source sizes and fluxes. For comparison, the shear detection significance is 11$\sigma$ on the same selection of $44$ groups (A. Leauthaud 2011, private communication).\footnote[1]{The significance quoted for the shear does not take into account the full covariance matrix, as we have done for the magnification measurement. Therefore this shear significance might be a bit optimistic.}

To improve S/N in this measurement, we stack the lenses, consisting of $44$ massive X-ray-detected galaxy groups in the COSMOS 1.64 deg$^2$ field. We measure an optimally weighted cross-correlation between the X-ray groups and high-redshift LBGs, with $1 \sigma$ error bars determined from jackknife resampling of the sources. Performing composite-halo fits to this optimally weighted signal yields a measurement of the relative scaling between shear- and magnification-derived masses. Our magnification measurement yields a mass $M_{\text{mag}}=aM_{\text{shear}}$ where the best-fit parameter $a= 1.2 \pm 0.4 \pm 0.4^{\text{sys}}$ (SIS), and $a= 1.8 \pm 0.5 \pm 0.4^{\text{sys}}$ (NFW), demonstrating a rough consistency with the shear measurement.

As discussed in Section 4.1, a central issue is the importance of having correct $(\alpha-1)$ measures for every source galaxy, to ensure that the {\it optimal weighting} truly is optimal. We perform a thorough error analysis that includes measurement uncertainties from the full covariance matrix, and investigate possible sources of systematic errors from both photometry and externally calibrated LF parameters.

We claim that LBGs are a preferred source sample when it comes to performing lensing magnification analyses using source number counts. A few reasons for the superiority of the LBG sample include more reliable redshift determinations, as well as greater lensing efficiencies and generally much higher values of the quantity $(\alpha-1)$. The single most significant reason to choose LBGs for this type of analysis, however, is for the ease of obtaining a reliable measure of $\alpha(m)$. Previous deep measurements of LBG LFs allow us to perform calculations yielding the optimal weight factor $(\alpha-1)$, as well as its associated uncertainty.

Although the S/N of shear is superior to magnification in general, the latter probes the surface mass density of the lens directly, while the shear measures the differential mass density. Thus the combination of these two independent measurements is desirable, and breaks the lens mass-sheet degeneracy. In fact, \citet{RozoSchmidt10} demonstrated that joining magnification into shear analyses, independent of survey details, can improve statistical precision by up to 40\%-50\%. Magnification using source number densities is also far less sensitive to the effects of atmospheric seeing than either shear or magnification using source sizes. Both of these methods require quality source images which, for very high redshift sources, can currently only be obtained from space-based data.

Improving the overall weak-lensing-derived constraints on cosmological and astrophysical parameters is not the only benefit to incorporating magnification into our analyses, however. Measurements of magnification are sensitive to completely different systematics than shear, and therefore uniquely positioned to help improve calibration of these residual effects on shear measurements. For example, magnification (using number counts) is not at all sensitive to the possible intrinsic alignment of source galaxies, since it does not use any shape information. Magnification can also be used as a simultaneous probe of intergalactic dust extinction, a small but measurable effect through its wavelength dependence \citep{Menard10}, and as a direct way to measure galaxy bias \citep{Waerbeke10}.

As one proceeds to investigate dark matter structures at increasingly high redshift, it becomes more and more important to include the magnification component of the signal. This is a direct consequence of the fact that higher redshift lenses necessitate more distant sources, which are in turn much harder to measure shapes for. Proceeding exclusively with shear necessarily means that a high fraction of detected sources are being eliminated from the lensing analysis, and information is therefore being lost, simply because we lack the capabilities to robustly determine their shapes. With photometric redshifts available, the possibility to do magnification studies on our shear catalogs really comes along free of charge. Upcoming projects will survey the entire extragalactic sky, and the inclusion of magnification will be a necessary component of any robust weak lensing study.

\vspace{10.pt}

The authors thank Fabian Schmidt and Martha Milkeraitis for useful discussions related to this work. J.F. was supported by JPL grant number 1394704, and is now supported by NSERC and CIfAR. H.H. is supported by the Marie Curie IOF 252760 and by a CITA National Fellowship. This work was performed in part at JPL, run by Caltech under a contract for NASA. This work was supported by World Premier International Research Center Initiative (WPI Initiative), MEXT, Japan. This work is based in part on data collected at Subaru Telescope, which is operated by the National Astronomical Observatory of Japan, and on observations made with the NASA/ESA {\it Hubble Space Telescope}. This research has made use of the NASA/IPAC Infrared Science Archive, which is operated by the Jet Propulsion Laboratory, California Institute of Technology, under contract with the National Aeronautics and Space Administration. This work is also based on observations obtained with MegaPrime/MegaCam, a joint project of CFHT and CEA/DAPNIA, at the Canada-France-Hawaii Telescope (CFHT) which is operated by the National Research Council (NRC) of Canada, the Institute National des Sciences de l'Univers of the Centre National de la Recherche Scientifique of France, and the University of Hawaii.



\end{document}